\documentclass[10pt,twocolumn,letterpaper]{article}

\usepackage[pagenumbers]{cvpr} %

\usepackage{ulem}
\newcommand\ourmethod{\textsc{TriTex}}

\definecolor{cvprblue}{rgb}{0.21,0.49,0.74}
\usepackage[pagebackref,breaklinks,colorlinks,allcolors=cvprblue]{hyperref}
\usepackage{xcolor}
\usepackage{array}
\usepackage{colortbl} 
\usepackage{arydshln}
\usepackage{makecell}

\usepackage{tablefootnote} 
\def\paperID{5630} %
\def\confName{CVPR}
\def\confYear{2025}

\title{TriTex: Learning Texture from a Single Mesh via Triplane Semantic Features}

\author{
Dana Cohen-Bar$^{1,2}$ \hspace{6mm}  
Daniel Cohen-Or$^1$ \hspace{6mm}  
Gal Chechik$^2$ \hspace{6mm} 
Yoni Kasten$^2$ \\[4pt]
$^1$Tel Aviv University	 \hspace{6mm} $^2$NVIDIA 
\\[-20pt]
}

\begin{document}
\twocolumn[{%
    \renewcommand\twocolumn[1][]{#1}%
    \maketitle
    \begin{center}
    \vspace{-8pt}
        \makebox[\textwidth][c]{%
            \includegraphics[width=1\textwidth]{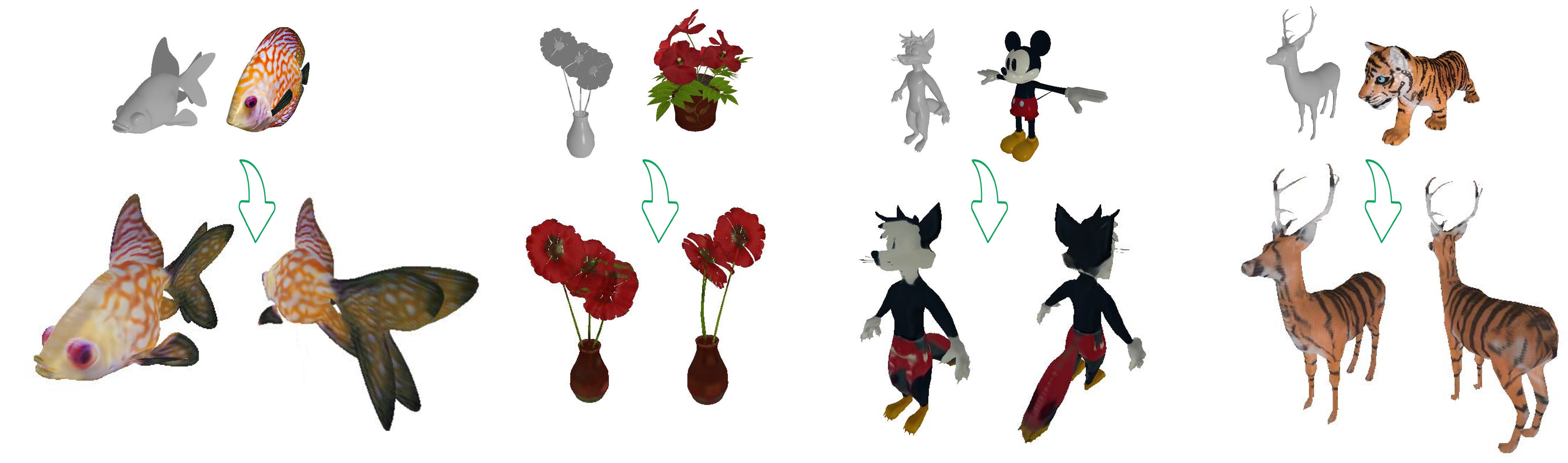}%
        }
        \captionof{figure}{
            TriTex is a method for learning to transfer a texture from a single mesh in a feed-forward manner to other input meshes. The figure shows four examples. In each, given a target geometry (up left) and a source textured mesh (up right) the texture is transferred to the target 3D object (bottom two views of the target object).
        }
    \label{fig:teaser}
    \end{center}
}]

\begin{abstract}
As 3D content creation continues to grow, transferring semantic textures between 3D meshes remains a significant challenge in computer graphics. While recent methods leverage text-to-image diffusion models for texturing, they often struggle to preserve the appearance of the source texture during texture transfer. We present \ourmethod, a novel approach that learns a volumetric texture field from a single textured mesh by mapping semantic features to surface colors. Using an efficient triplane-based architecture, our method enables semantic-aware texture transfer to a novel target mesh. Despite training on just one example, it generalizes effectively to diverse shapes within the same category. Extensive evaluation on our newly created benchmark dataset shows that \ourmethod{} achieves superior texture transfer quality and fast inference times compared to existing methods. Our approach advances single-example texture transfer, providing a practical solution for maintaining visual coherence across related 3D models in applications like game development and simulation.\\
Project page: \href{https://danacohen95.github.io/TriTex/}{ https://danacohen95.github.io/TriTex/.}  
\end{abstract}

\section{Introduction}
\label{sec:intro}
Texturing 3D objects is a fundamental task with wide-ranging applications in game development, simulation, and video production. For example, when generating 3D scenes, style needs to be consistently applied across multiple 3D models that share semantic properties but differ in shape, such as in environments featuring diverse buildings or plants. Improving texture transfer can streamline texturing workflows by maintaining visual coherence across related models and preserving texture details even during mesh modifications, ensuring a consistent appearance across all elements. With the advent of large generative models, new approaches to 3D texturing have emerged, moving beyond traditional procedural methods to enable automatic texturing and texture transfer \cite{texturify,Auv-net}. However, these techniques often require the availability of large databases of 3D objects and extensive training for each class to achieve high-quality results. 

Recent texturing methods use text-to-image diffusion models to texture meshes without requiring 3D data collection. Some methods~\cite{latentnerf,Paint3D,TextureDreamer} apply an SDS loss to optimize the texture map, which tends to be a slow process. Others~\cite{TEXTure,TexFusion, Text2Tex, syncmvd, SyncTweedies,mvedit} employ iterative depth-conditioned image inpainting on rendered mesh views, synchronizing across views to maintain consistency. However, when performing texture transfer, these methods struggle to faithfully preserve the appearance of the source texture. 

We introduce \ourmethod, a novel technique that learns a volumetric texture field from a single textured mesh,\hspace{10pt} mapping its semantic features to RGB colors on its surface. During inference, given a target mesh, this learned field enables texture transfer by predicting colors based on the target mesh's semantic features, allowing the source texture to be applied in a way that maintains semantic correspondences between the source and target meshes (see Figure~\ref{fig:teaser}). Despite being trained on a single object, our trained volumetric texture generalizes well to novel objects within the same category. Our fast inference time enables the stylization of large scenes 
More specifically, to learn this texture field, we first extract semantic features using a frozen Diff3F~\cite{Diff3F} model, then project these features onto a triplane, and process them through a mapping network to produce the final triplane features. 

Our experiments demonstrate the effectiveness of our texture transfer method, showing superior quality compared to previous approaches while maintaining fast inference times. We evaluate our approach using automated metrics and human evaluation on a benchmark dataset that we created specifically for the texture transfer task. Our method shows superior performance compared to previous approaches.

To summarize, our contributions are: (1) Introducing a method for transferring textures from a single textured mesh to new shapes while preserving semantic relations. (2)~Leveraging pre-trained 3D semantic features, reprojecting them into a triplane representation, and enabling effective processing through convolutional layers. (3) Demonstrating strong generalization to novel objects with significant shape variations, despite being trained on only a single example. (4) Achieving high-quality texture transfer with fast inference, outperforming previous methods in both speed and quality.

\section{Related Work}
\begin{figure*}[ht!]
    \centering
    \includegraphics[width=0.95\linewidth,trim={0cm 0cm 0cm 0cm}, clip]{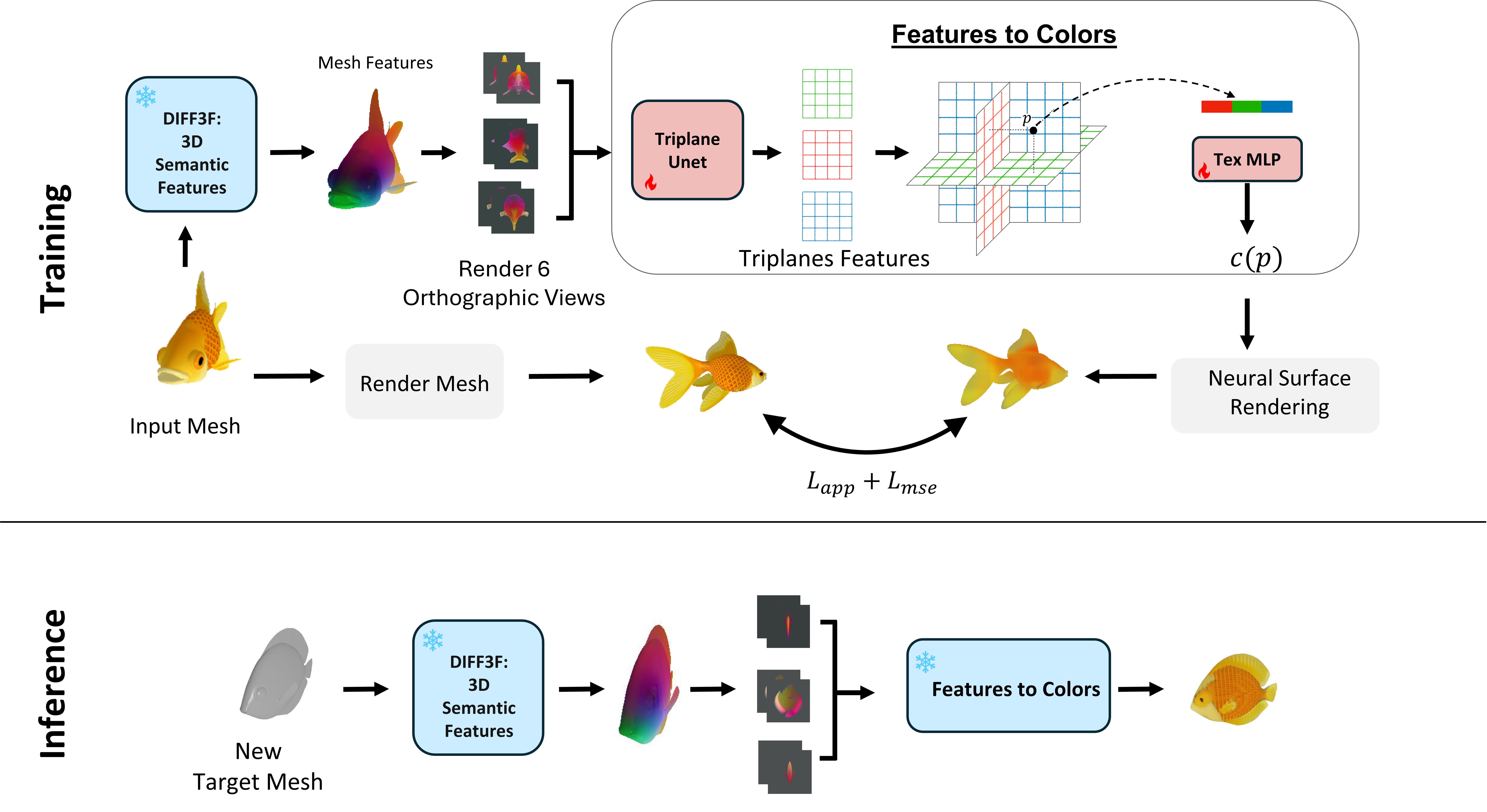}

    \caption{
\textbf{Training Pipeline (Top). } Given an input textured mesh and its pre-extracted Diff3F features, we project six orthographic views to create an initial triplane. This triplane is processed using triplane-aware convolutional blocks, which, along with the texture MLP, define a coloring neural field. This field, together with the input geometry, is used to render the colored mesh. Appearance losses are applied between the true mesh appearance and the rendered appearance.  
\textbf{Inference (Bottom).} Given a new mesh (left), our pre-trained model maps its semantic properties to colors (right), transferring the texture from the original textured mesh learned during the training phase. }
    \label{fig:method}
       \vspace{-8pt}
\end{figure*}

\paragraph{Texture Synthesis}
Classical methods use sampling approaches to generate 2D textures \cite{de1997multiresolution, efros1999texture, heeger1995pyramid, zhu1998filters, wei2001texture}. More recent approaches rely on deep learning to create textures based on training data \cite{xian2018texturegan, zhou2018non}. Texturify \cite{texturify} introduced a Generative Adversarial Network (GAN) for texturing 3D models using an example image, applying face convolutional operators directly on the 3D object’s surface. Auv-net \cite{Auv-net} predicts UV mapping and texture images from 3D geometry, ensuring consistent alignment across elements within the same category, which enables effective training of generative texture models with this embedding. Both Auv-net and Texturify require large datasets of predefined categories to train their generative models, whereas our approach is trained solely on a single textured mesh and can generalize to a variety of objects with similar semantics.

\paragraph{Leveraging Large Vision-Language Pretrained Models for Texture Generation }
Recent advances in image-language foundation models have enabled effective texture transfer and synthesis without requiring additional training data. Text2Mesh \cite{Text2Mesh} and TANGO \cite{TANGO} edit 3D mesh appearance by aligning rendered images with a text prompt in CLIP space. Dream3D \cite{Dream3D} optimizes the appearance of generated geometry using a Neural Radiance Field (NeRF) by applying a CLIP loss. Latent-NeRF \cite{latentnerf} applies the SDS-loss \cite{poole2022dreamfusion} to rendered images to optimize a NeRF representation in the stable diffusion latent space. More recently, Paint-it \cite{paintit}, Fantasia3D \cite{fantasia3d}, and FlashTex \cite{FlashTex} have enhanced SDS-loss-based texturing by incorporating Physically Based Rendering (PBR), BRDF modeling, and illumination control, respectively. These optimization techniques require processing each object individually, leading to slower performance, whereas our method has a much shorter inference time.

Other methods employed optimization-free approaches by directly applying depth-conditioned diffusion models in sequence. Texture~\cite{TEXTure} applies depth-conditioned inpainting on sequential mesh projections to achieve mesh texturing. Concurrent and follow-up methods \cite{TexFusion, Text2Tex, syncmvd, SyncTweedies,mvedit} introduced improvements, with Text2Tex\cite{Text2Tex} proposing an automatic per-object view selection scheme, and TexFusion~\cite{TexFusion}, SyncMVD~\cite{syncmvd}, SyncTweedies~\cite{SyncTweedies} and MVEdit \cite{mvedit} synchronizing views by operating on intermediate denoising steps. Paint3D~\cite{Paint3D} additionally trains a model to remove illumination effects from the generated texture map, enabling relighting of the textured meshes. In~\cite{Meta3dTextureGen}, a geometry-aware multiview diffusion model is used to enhance view consistency and efficiency.

\paragraph{Texture Transfer}
The problem of texture transfer has always been an area of interest and research, with early work aiming to transfer both stochastic texture  \cite{Context_aware_textures} and semantic texture \cite{mesh_match, geometry_correlation} . 
More recently, several methods leverage 3D data for texture transfer tasks. 
3DStyleNet \cite{3DStyleNet} transforms a source mesh by applying part-aware low-frequency deformations and generating texture maps, using a target mesh as a style reference.
Mesh2Tex~\cite{Mesh2Tex} learns a texture manifold from a large dataset of 3D objects and images. This enables new images to be projected into the manifold and transferred to 3D objects in trained categories, such as cars and chairs. Personalization techniques for diffusion models~\cite{ruiz2023dreambooth,gal2022image} are used in methods like TEXTure and TextureDreamer~\cite{TEXTure,TextureDreamer} to texture a mesh from a few input images, which can be either natural images or rendered images from another textured mesh. MVEdit~\cite{mvedit}, EASI-Tex~\cite{EASI-Tex} and Paint3D \cite{Paint3D} transfer textures from an image to a 3D object using IP-Adapter~\cite{ye2023ip}, where EASI-Tex further enhances geometry details by incorporating a rendered mesh edge map as an additional control for image generation. While these methods are efficient, they often fail to faithfully preserve the appearance of the source texture. In contrast, our method maintains both efficiency and accurate texture preservation.

Auv-net~\cite{Auv-net} enables texture transfer between meshes through a learned UV mapping and aligned texture representations. Mitchel \etal \cite{fieldlatents} proposed a diffusion process within a learned latent space on the surface, allowing the transferring of textures to other meshes using an input semantic label map. Nerf Analogies~\cite{fischer2024nerfanalogiesexamplebasedvisual} extends texture transfer to neural radiance fields, enabling appearance transfer between source and target NeRFs with semantic similarities, though it requires per-pair training for each set of NeRFs.

\paragraph{Semantic Features and Correspondences}
Finding corresponding points between images has traditionally relied on local feature descriptors~\cite{lowe2004distinctive} that are invariant to lighting and color. Later methods used features extracted from pre-trained classification models~\cite{aberman2018neural,simo2015discriminative}, providing invariance to local shape deformations and leveraging higher-level semantic relationships. Recently, Amir \etal \cite{amir2021deep} demonstrated the effectiveness of features from the pre-trained Vision Transformer DINO-ViT~\cite{caron2021emerging} as semantic descriptors. Splice~\cite{tumanyan2022splicing} uses DINO-ViT features to transfer colors from a source image to a target image structure by preserving the target’s local DINO features while adapting its global appearance to match the source. Neural Congealing~\cite{ofri2023neural} employs DINO-ViT semantic features to map a set of semantically related images into a shared canonical representation, enabling zero-shot joint editing. More recently, DIFT~\cite{tang2023emergent} introduced semantic features extracted from a pre-trained diffusion model during denoising. Diff3F~\cite{Diff3F} mapped both DIFT and DINO features onto a 3D mesh, resulting in 3D semantic features that enable shape correspondences.
\hspace{15pt} Our model benefits from these recent advances in semantic features and uses them to train a texture transfer model with a single textured mesh.

\paragraph{Models Trained on a Single Instance} Several methods have been developed to train neural networks using a single example. SinGAN ~\cite{SinGAN} trains a model to capture the internal distribution of patches within a single image, enabling the generation of diverse samples with similar visual content. Similarly, Wu \etal \cite{3D_Shapes_from_Single_Example} and Hertz \etal  \cite{Deep_geometric_texture_synthesis} use GANs to generate shape and geometric texture variations, respectively, from a single mesh. More recent works~\cite{wang2022sindiffusion, kulikov2023sinddm, nikankin2022sinfusion} have employed diffusion models trained on individual images to produce image variations, with Sinfusion~\cite{nikankin2022sinfusion} also extending this approach to single videos to create video variations. Additionally, Sin3DM~\cite{Sin3DM} and~\cite{fieldlatents} apply similar techniques to generate variations of a single textured mesh, either for textured mesh variations~\cite{Sin3DM} or for texture variations~\cite{fieldlatents}.
Similarly, we propose training a mesh texturing model using just a single textured mesh.

\paragraph{Triplane Representation} The triplane representation introduced in EG3D~\cite{EG3D} consists of three 2D feature grids aligned with three distinct orthogonal planes in the 3D coordinate system. This setup enables image operations, such as convolutional neural layers, and uses interpolation to define a continuous feature grid suitable for neural fields. Triplane representation has been applied in 3D-aware GANs~\cite{EG3D} and in mesh generation using diffusion models~\cite{3DGen,3dneuralfieldgeneration}. For a detailed technical description, see Sec.~\ref{section::background}.

\section{Method}

\subsection{Background}
\label{section::background}
\paragraph{Triplane Representation}
The triplane representation, introduced in EG3D~\cite{EG3D}, encodes 3D information using three axis-aligned orthogonal feature planes \(F_{XY}, F_{XZ}, F_{YZ}\), each of size \(\mathbb{R}^{W \times H \times C}\), where \(W\) and \(H\) represent the spatial resolution, and \(C\) is the number of channels. To query features at any 3D position \(x \in \mathbb{R}^3\), the coordinates are used to sample from each of the three feature planes (XY, XZ, YZ) via bilinear interpolation, yielding three feature vectors \(f_{xy}, f_{xz}, f_{yz}\in\mathbb{R}^C\). These feature vectors are then concatenated and passed through a lightweight MLP decoder network to produce the final output. This approach enables efficient feature extraction while maintaining expressiveness through learned feature integration, offering a balance between efficiency and quality.

Importantly, the triplane representation enables processing with standard 2D convolutional layers, which are far more efficient than using 3D convolutional layers on volumetric grids. In EG3D ~\cite{EG3D}, 2D convolutional layers were applied on the three concatenated feature planes. Recently, Sin3DM ~\cite{Sin3DM} introduced the triplane-aware convolution block, suggesting a more geometrically integrated approach. Instead of processing the three feature grids (\(F_{XY}, F_{XZ}, F_{YZ}\)) independently or simply concatenating them, this method aggregates features from each plane into the others. Let \(F_{XY}', F_{XZ}', F_{YZ}'\) represent the output of a single independent convolutional layer. Each plane is averaged over its axes and replicated along the third axis, before being concatenated into the relevant plane. For example, \(F_{XY}'\) is averaged over the \(x\) and \(y\) axes, producing \(F_{Y}'\) and \(F_{X}'\) respectively. These are then replicated along the \(z\)-axis and concatenated with \(F_{YZ}'\) and \(F_{XZ}'\), respectively.

\paragraph{Diff3f Semantic Features}
Our method leverages DIff3F \cite{Diff3F} features, which provide robust semantic descriptors for 3D shapes without requiring textured inputs or additional training. DIff3F extracts semantic features by utilizing foundational vision models in a zero-shot manner. The process begins by rendering depth and normal maps from multiple views of the input mesh. These maps serve as conditioning inputs to ControlNet \cite{controlnet} and Stable Diffusion \cite{stablediffusion} for generating view-dependent features.  The features from the diffusion process maintain semantic consistency despite potential\hspace{2pt} visual\hspace{2pt} variations\hspace{2pt} across\hspace{2pt} views, and are aggregated onto the original 3D surface. These features, further enriched with DINO features \cite{caron2021emerging}, capture rich semantic information and enable reliable correspondence across shapes with significant geometric variations, making them well-suited for our texture transfer task.

\begin{figure}[!htbp]
    \begin{tabular}{@{\hspace{1em}}p{0.1\textwidth}@{\hspace{5em}}p{0.4\textwidth}@{}}
        Source & Texture Transfer Results \\  
    \end{tabular}

    \includegraphics[width=0.48\textwidth]{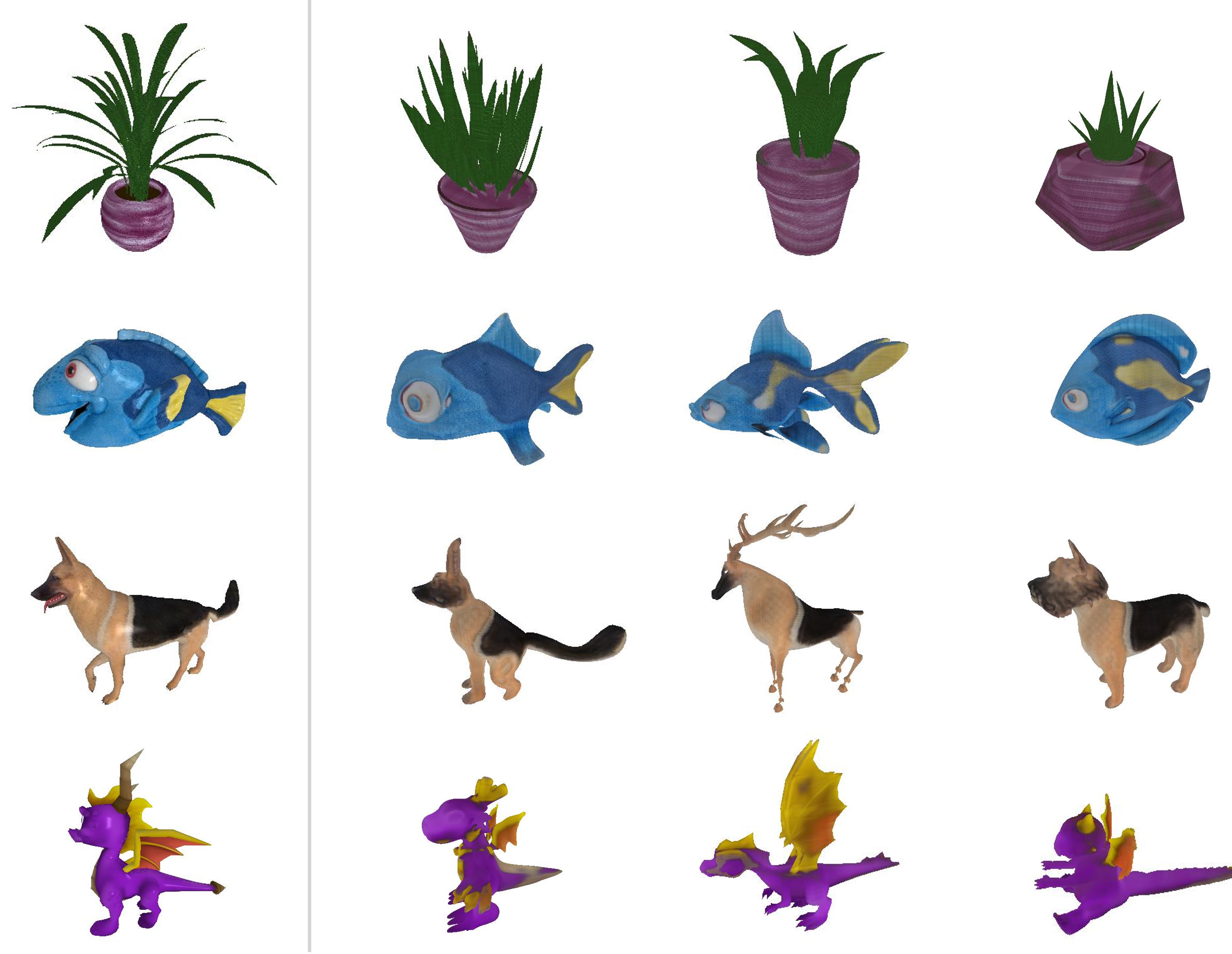}
    \caption{\textbf{Texture Transfer Results.} Given a source mesh (left column), our network transfers its appearance onto target meshes of varying shapes (three right columns). Each row demonstrates texture transfer using a different source mesh. }
    \label{fig:packs}
\vspace{-15pt}
\end{figure}

\subsection{Our approach: \ourmethod}
Given a textured source mesh, our goal is to train a feed-forward texturing function that can transfer that texture to any target mesh while preserving the semantics of the texture. 
Our pipeline is shown in \cref{fig:method}.

The architecture, based on a deep neural network, processes a 3D mesh with pre-extracted semantic features and a query 3D point, and outputs the corresponding color for the query point. Formally, we define a learnable function: \( f(M, \mathbb{R}^3) \rightarrow [0,1]^3 \), where \( M = (V, S, E, F) \) is the input mesh, with \( V = \{v_1, \dots, v_n \mid v_i \in \mathbb{R}^3 \} \) and \( S = \{s_1, \dots, s_n \mid s_i \in \mathbb{R}^D \} \), representing the 3D vertices and their corresponding semantic features, pre-extracted by ~\cite{Diff3F}. Here, \( E \) and \( F \) represent the edges and faces, respectively, defined on the vertices.

The first part of our architecture processes \( M \) by orthographically projecting the semantic features into an axis-aligned feature triplane \( \mathcal{T} \in \mathbb{R}^{3 \times W \times H \times 2D} \), where each feature plane is created by concatenating features from two orthographic projections taken from opposite directions. To address the low feature resolution (32×32)  and enable fine detail generation, we incorporate positional encoding into the input. Triplane-aware convolutional 2D blocks \cite{Sin3DM} (see further details in Sec.\ref{section::background}) are then applied to the triplane to produce a modified triplane \( \mathcal{T}' \in \mathbb{R}^{3 \times W \times H \times D'} \). Finally, to predict the color for a query 3D point, we sample features from the processed planes in \( \mathcal{T}' \) according to the query point's location, concatenate them to form a single feature vector, and pass this vector through the coloring MLP \( c: \mathbb{R}^{3D'} \rightarrow [0,1]^3 \), which maps it to the final RGB color for the point.

\paragraph{Training}

We train our model on a single textured mesh using a rendering-based reconstruction loss, comparing rendered views of our predicted textures with ground truth views of the source mesh from randomly sampled camera angles. Let \( I_R(\theta) \) be an image rendered by our pipeline from angle \( \theta \), with the corresponding colors predicted by the MLP \( c \), generated by querying 3D points on the mesh through intersecting rendered camera rays. This image is conditioned only on the input semantic features defined for the geometry and depends on the differentiable parameters of the triplane-aware convolutional layers and \( c \). To train the learnable parameters, we use the ground truth rendered images, which are known for the single training textured mesh, and apply the MSE loss on the image pixels:
\[
\mathcal{L}_{\text{MSE}}(\theta) = \frac{1}{N} \sum_{i=1}^{N} \left\| I_R(\theta)_i - I_{\text{GT}}(\theta)_i \right\|^2,
\]
where \( I_{\text{GT}}(\theta) \) is the ground truth image, and \( i \) indexes the pixels in the image. While MSE loss encourages pixel-level accuracy, it may not effectively capture high-level perceptual details. To address this, we incorporate a perceptual loss \( \mathcal{L}_{\text{app}} \) from Tumanyan \etal ~\cite{splicing}, which emphasizes high-level semantic features. This improves the alignment of these features between the generated and ground truth images, enhancing texture realism. 
Our final loss function is then:
\[
    \mathcal{L} = \mathbb{E}_{\theta} \left[ \mathcal{L}_{\text{MSE}}(\theta) + \delta_{\text{app}} \mathcal{L}_{\text{app}}(\theta) \right],
\]
where \( \mathbb{E}_{\theta} \) represents the expectation over all sampled camera angles, and \( \delta_{\text{app}} \) controls the relative weight of \( \mathcal{L}_{\text{app}} \) compared to \( \mathcal{L}_{\text{MSE}} \). To improve generalization, we apply two levels of augmentation: (1) preprocessing augmentation, where simple 3D transformations are applied to the input mesh and features are extracted for each variant to enrich the learned feature distribution, and (2) training-time augmentation, where translation, scaling, and small rotational perturbations are applied to the mesh during training.

\begin{figure}[!htbp]
    \centering
    \setlength{\tabcolsep}{1pt}
    {\small
    \begin{tabular}{ccc}
        \multicolumn{3}{c}{} \\
        \includegraphics[width=0.32\linewidth, trim=60 60 20 160, clip]{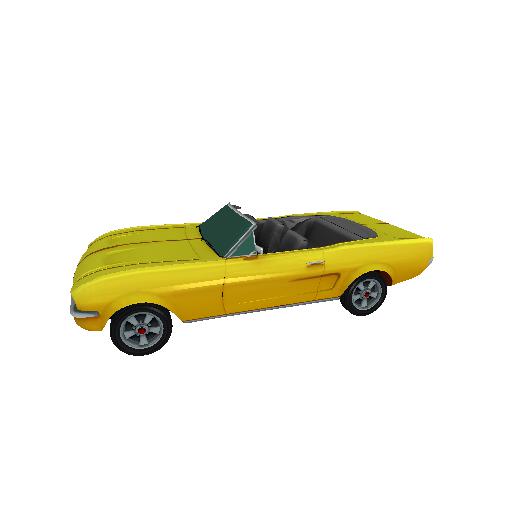} &
        \includegraphics[width=0.32\linewidth, trim=60 60 20 160, clip]{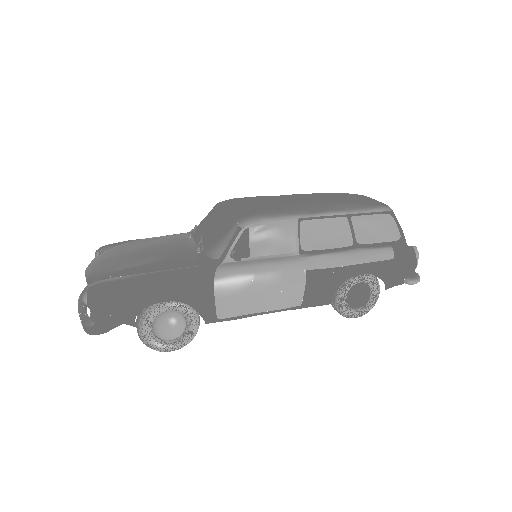} &
        \includegraphics[width=0.32\linewidth, trim=60 60 20 160, clip]{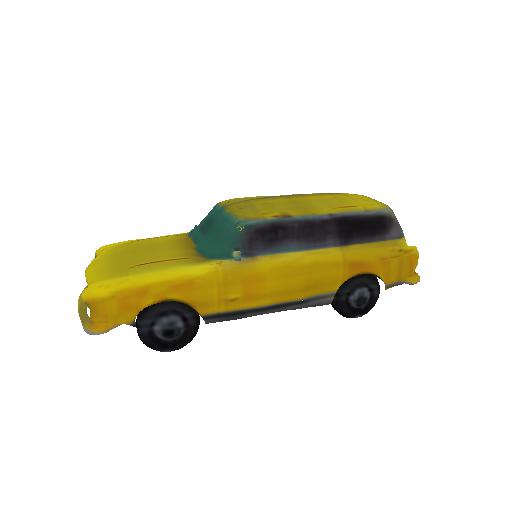} \\[-10pt]
        \includegraphics[width=0.32\linewidth, trim=60 60 20 70, clip]{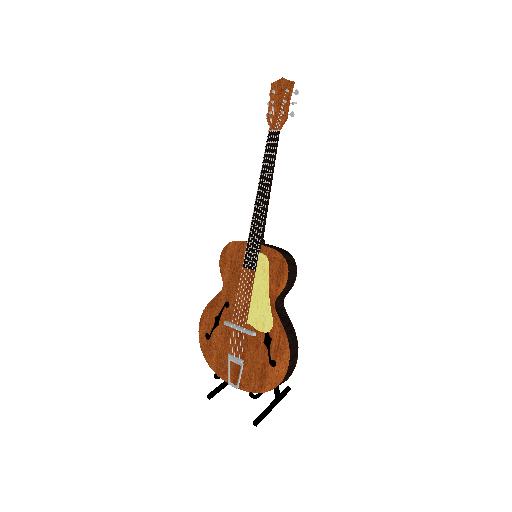} &
        \includegraphics[width=0.32\linewidth, trim=60 60 20 70, clip]{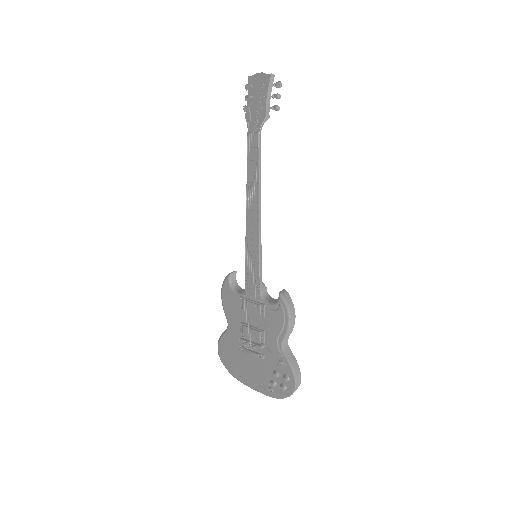} &
        \includegraphics[width=0.32\linewidth, trim=60 60 20 70, clip]{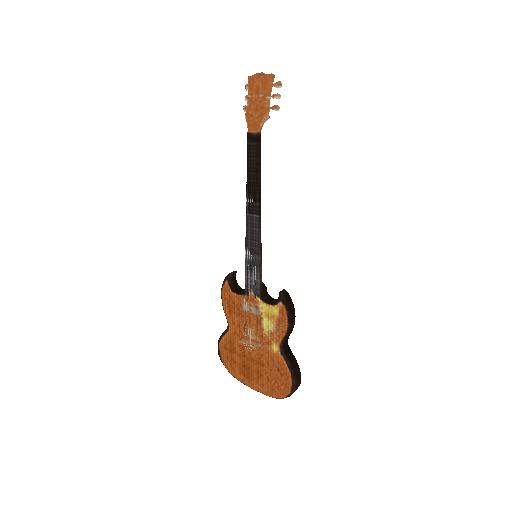} \\[-10pt]
        \includegraphics[width=0.32\linewidth, trim=60 60 20 70, clip]{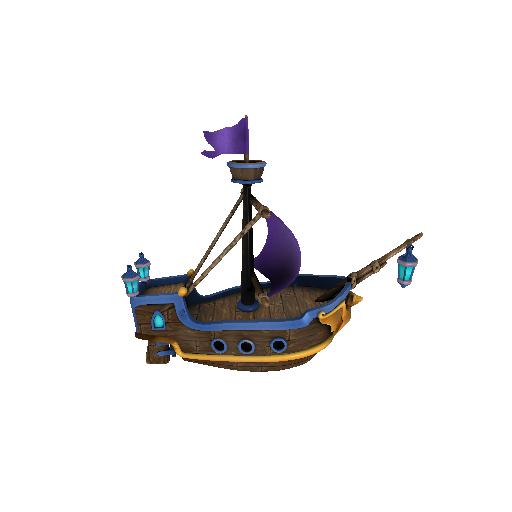} &
        \includegraphics[width=0.32\linewidth, trim=60 60 20 70, clip]{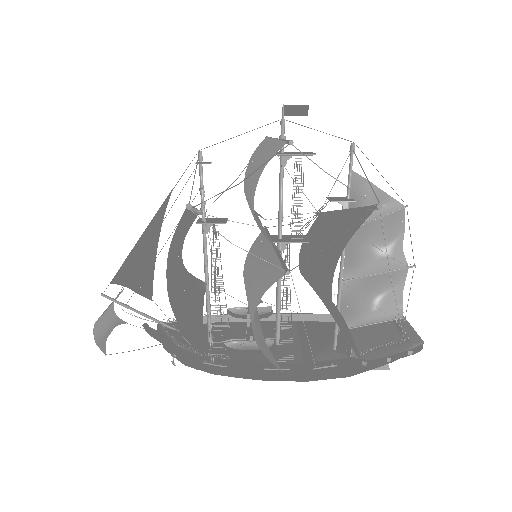} &
        \includegraphics[width=0.32\linewidth, trim=60 60 20 70, clip]{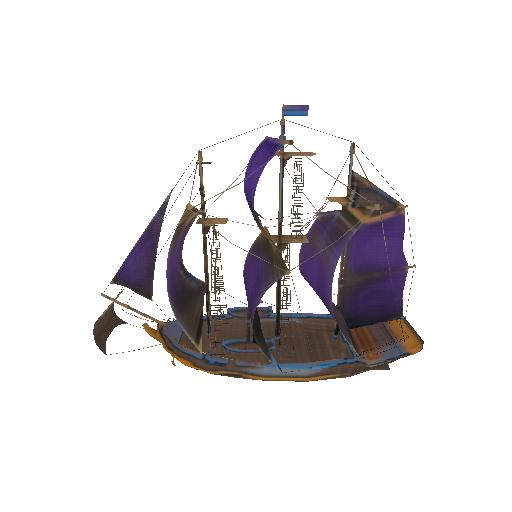} \\[-10pt]
        
        \includegraphics[width=0.32\linewidth, trim=60 60 20 80, clip]{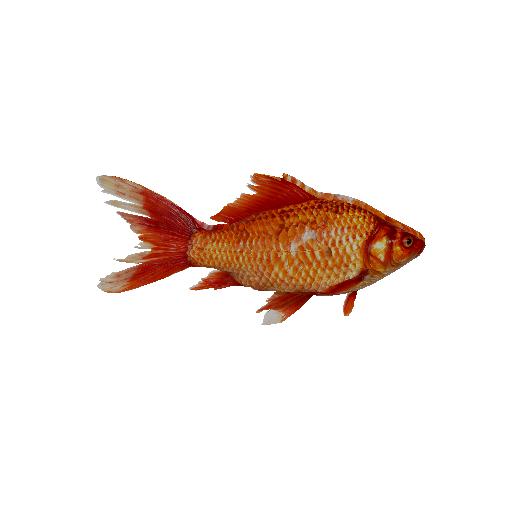} &
        \includegraphics[width=0.32\linewidth, trim=60 60 20 80, clip]{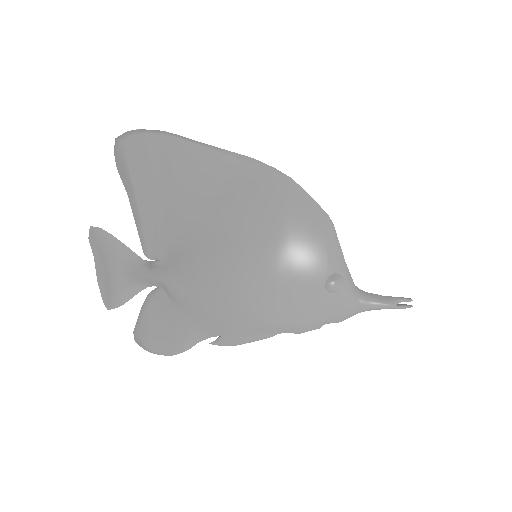} &
        \includegraphics[width=0.29\linewidth, trim=0 0 0 20, clip]{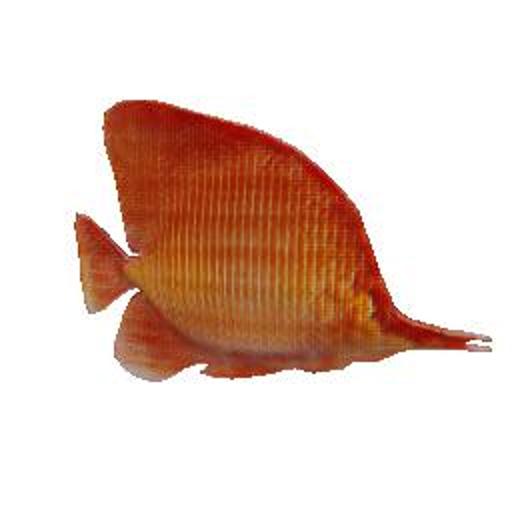} \\[-10pt]
        
        Source-textured & Target geometry & Target-textured \\
    \end{tabular}
    }
    \vspace{-8pt}
    \caption{
    \textbf{Additional Qualitative Results. }Showing the target geometry and the texture transfer. 
    }

    \label{fig:results}
\end{figure}

\section{Experiments And Results}
\newlength{\compfigwidth}
\setlength{\compfigwidth}{0.38\linewidth}

\begin{figure*}
    \centering
    \setlength{\tabcolsep}{-4pt}  
    {\scriptsize
    \begin{tabular}{@{}ccccc@{}} 
        Source Shape  & TEXTure & EASI-tex & MVEdit & Ours (\ourmethod) \\
                
        \includegraphics[width=\compfigwidth,trim={2cm 2cm 1cm 0cm}, clip]
        {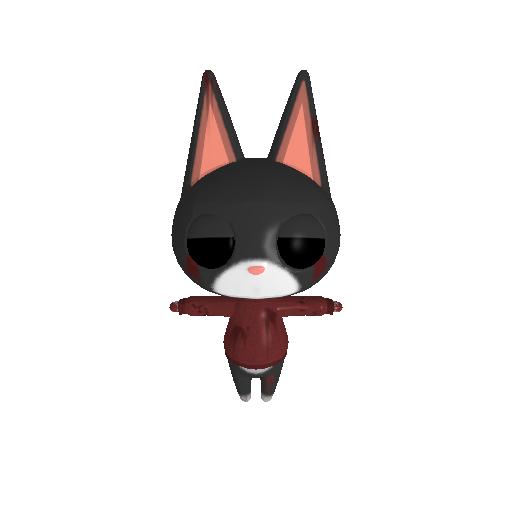} &

        \includegraphics[width=\compfigwidth, trim={0cm 0cm 0cm 0cm},clip]{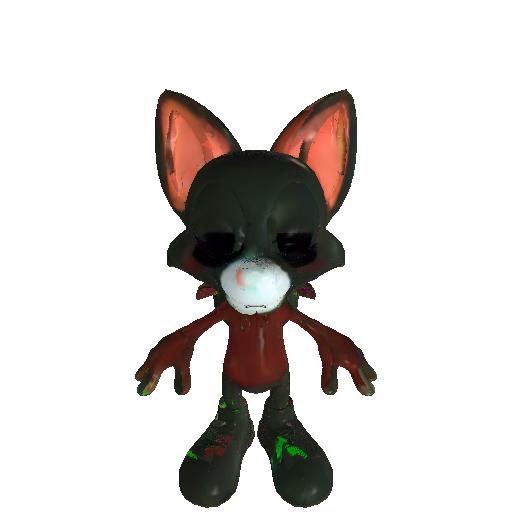} &
        \includegraphics[width=\compfigwidth, trim={0cm 0cm 0cm 0cm},clip]{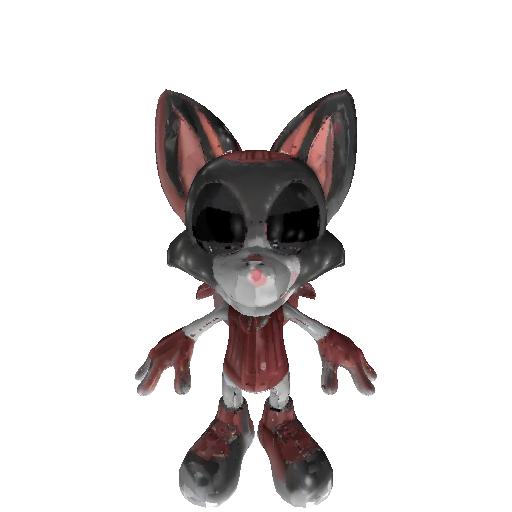} &
        \includegraphics[width=\compfigwidth, trim={0cm 0cm 0cm 0cm},clip]{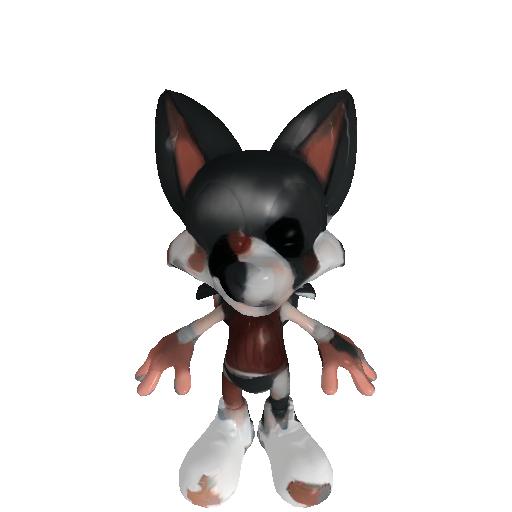} &
        \includegraphics[width=\compfigwidth, trim={0cm 0cm 0cm 0cm},clip]{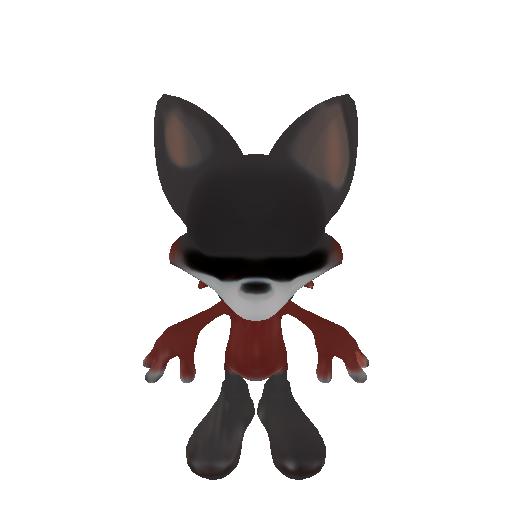} \\

        \includegraphics[width=\compfigwidth,trim={0cm 0cm 0cm 0cm}, clip]
        {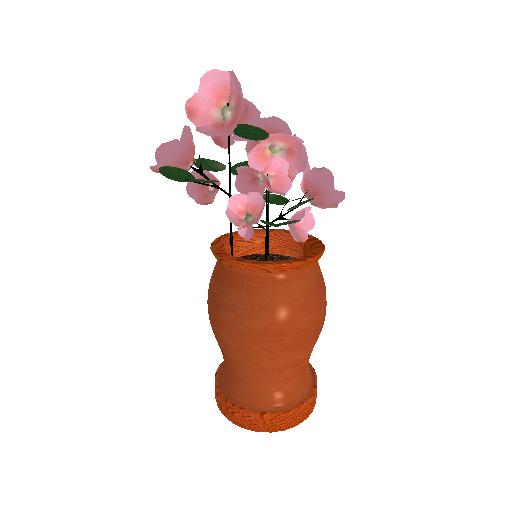} &
        \includegraphics[width=\compfigwidth,trim={2cm 2cm 2cm 2cm}, clip]{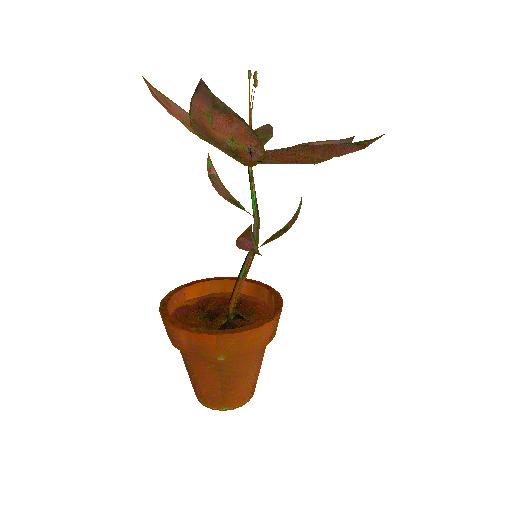} &
        \includegraphics[width=\compfigwidth,trim={2cm 2cm 2cm 2cm}, clip]{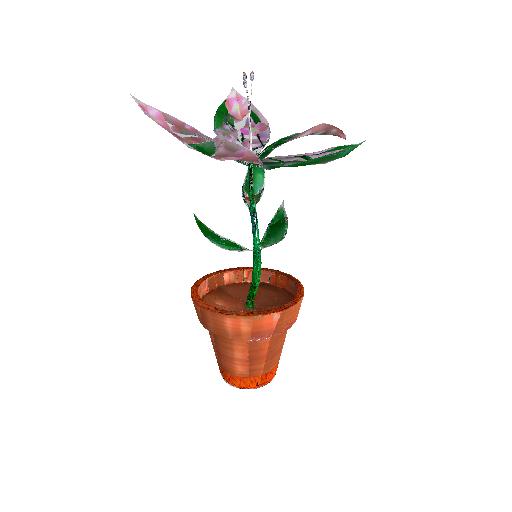} &
        \includegraphics[width=\compfigwidth,trim={2cm 2cm 2cm 2cm}, clip]{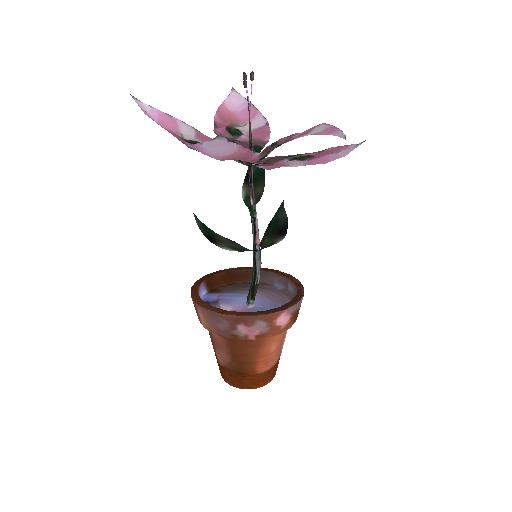} &
        \includegraphics[width=\compfigwidth,trim={2cm 2cm 2cm 2cm}, clip]{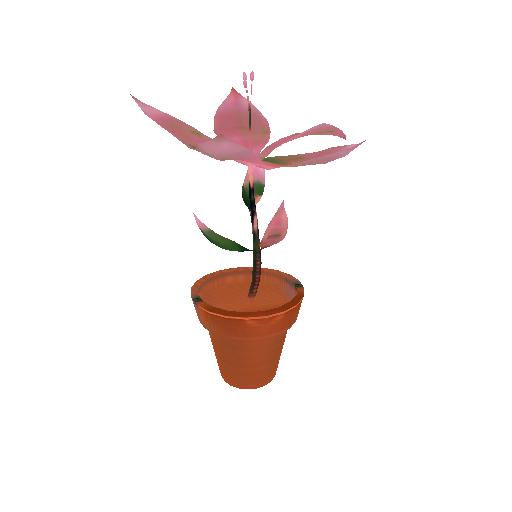}  \\[-10pt]
        
        \includegraphics[width=\compfigwidth,trim={0cm 2cm 2cm 3cm}, clip]{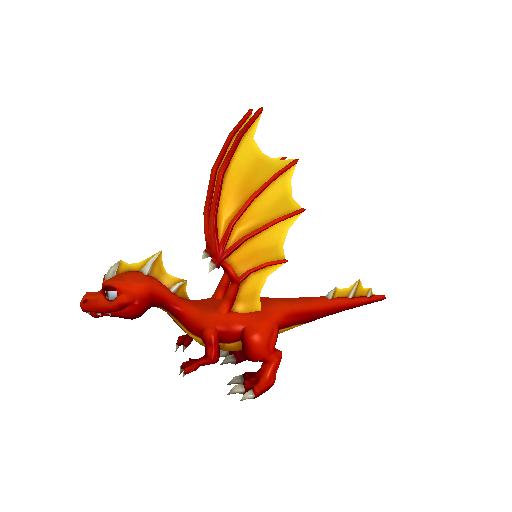} &
        \includegraphics[width=\compfigwidth,trim={0cm 0cm 0cm 0cm}, clip]{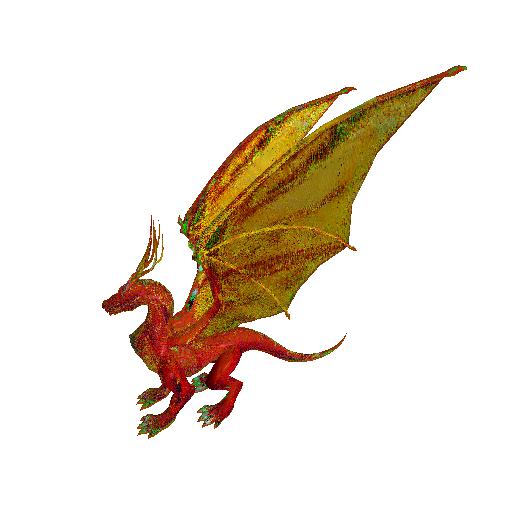} &
        \includegraphics[width=\compfigwidth,trim={0cm 0cm 0cm 0cm}, clip]{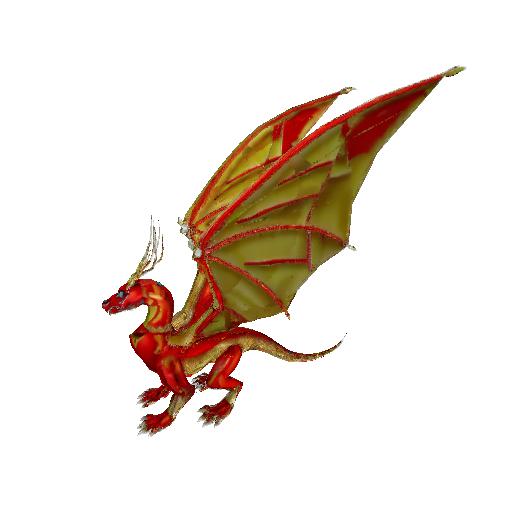} &
        \includegraphics[width=\compfigwidth,trim={0cm 0cm 0cm 0cm}, clip]{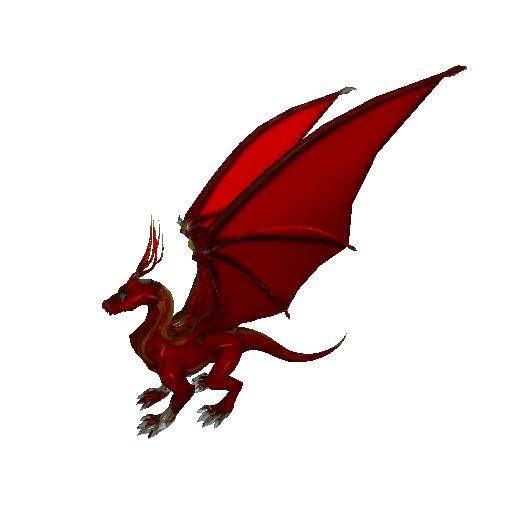} &
        \includegraphics[width=\compfigwidth,trim={0cm 0cm 0cm 0cm}, clip]{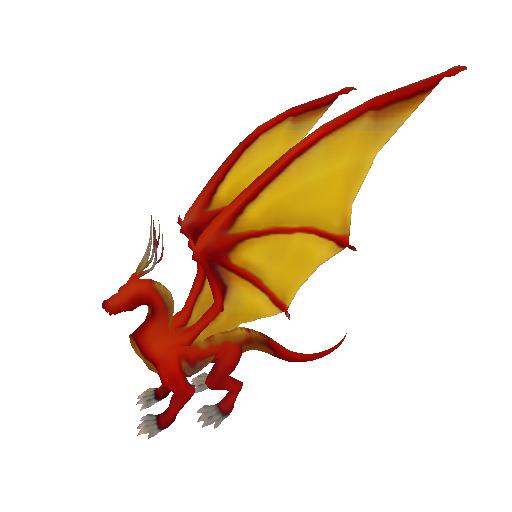} \\[-20pt]

        \includegraphics[width=\compfigwidth,,trim={1cm 1cm 2cm 2cm}, clip]{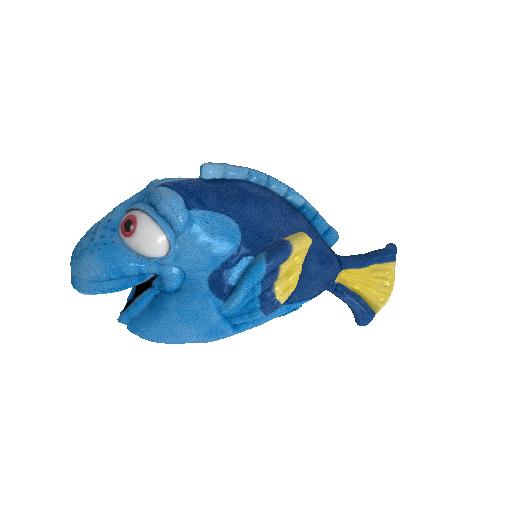} &
        \includegraphics[width=\compfigwidth]{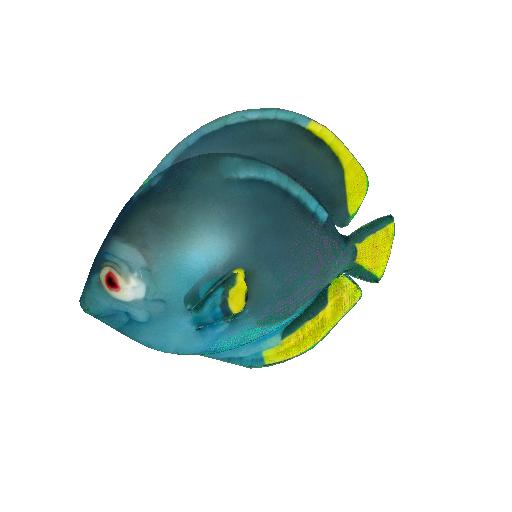} &
        \includegraphics[width=\compfigwidth]{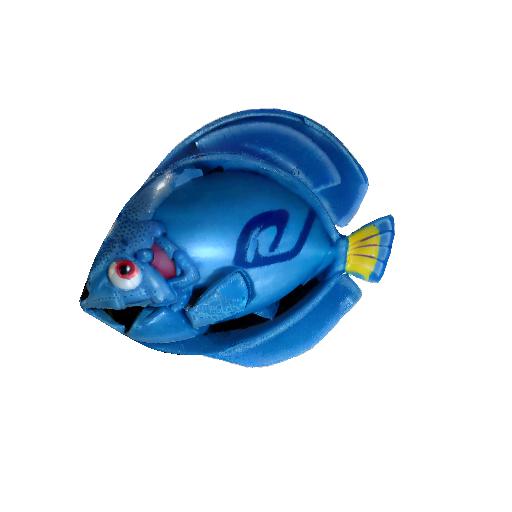} &
        \includegraphics[width=\compfigwidth]{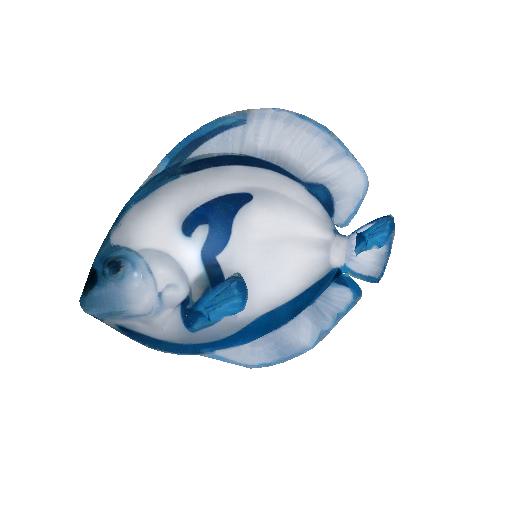} &
        \includegraphics[width=\compfigwidth]{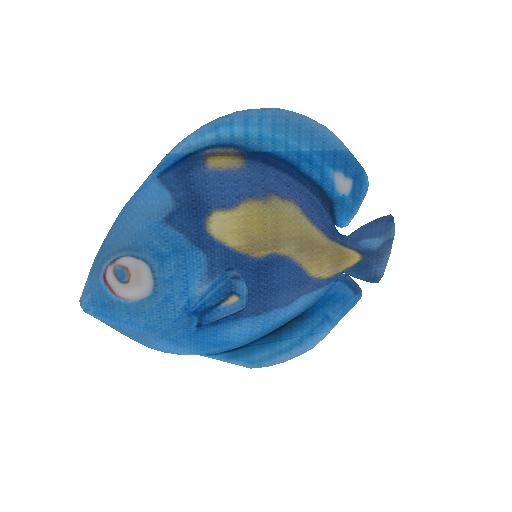} \\
        
    \end{tabular}
    }
    \vspace{-24pt}
    \caption{ \textbf{Qualitative Comparisons with baselines.} A comparison between \ourmethod{} (right-most column)  and state-of-the-art approaches for texture transfer. As observed, \ourmethod{} produces much more plausible texture transfer results.}
    \label{fig:comparisons}
\end{figure*}

\subsection{Experiment Details}

\paragraph{Dataset:} 
Our experimental dataset consists of 52 3D meshes across 9 categories, curated from Objaverse \cite{objaverse, ObjaverseXL}, a large-scale 3D object database. We first generated descriptive captions for each 3D object to identify semantically related objects using BLIP \cite{li2022blipbootstrappinglanguageimagepretraining}. We then used keyword filtering on these captions to group objects into similar categories. The final selection includes only objects that both share semantic similarities and maintain consistent orientations within their respective categories.
Overall, we have $256$ different pairs of source and target objects. The full list of source and target objects is provided in the supplemental.

\paragraph{Baseline Methods:}
We compare our method against three state-of-the-art optimization-free approaches for visually-guided mesh texturing. \textbf{(1) TEXTure} \cite{TEXTure}, a personalization-based approach, enabling texture transfer from a source mesh to a target mesh. \textbf{(2) EASI-TEX} \cite{EASI-Tex} and \textbf{(3) MVEdit }\cite{mvedit} use IP-Adapter to incorporate image guidance into their texture generation process, allowing direct control over the generated textures through reference images.

\begin{table}[h!]
    \caption{\textbf{Quantitative Evaluation.} Quantitative comparison of our method with baseline methods. We evaluate two appearance metrics and running time, showing that our method best preserves the source texture while achieving competitive speed.}
    \label{tab:performance_comparison}
    {\small
    \begin{tabular}{@{}lccc@{}}
        \toprule
        \textbf{Method} & \textbf{SIFID $\downarrow$} & \textbf{CLIP sim. $\uparrow$} & \textbf{Inference Runtime} \\ 
        \midrule
        TEXTure & 0.34 & 0.84 & 5 min. \\
        MVEdit & 0.38 & 0.84 &  1 min. \\
        EASI-TEX & 0.29 & 0.85 &  15 min. \\
        TriTex (ours) &  \textbf{0.22} & \textbf{0.87} & 1 min. \\
        \bottomrule
    \end{tabular}
    }
\end{table}

\paragraph{Metrics:}
We evaluate our method using human evaluation and two automated metrics: CLIP similarity score \cite{clip} and Single Image Frechet Inception Distance (SIFID). The CLIP score, adopted following TextureDreamer \cite{TextureDreamer}, measures the semantic and visual similarity between source and generated textures, indicating how well our method preserves the desired appearance. SIFID was originally proposed by \cite{SinGAN} to measure internal patch statistics within single images. Following Sin3DM \cite{Sin3DM}, we adopt this metric to evaluate texture similarity between source and target shapes.

For both metrics, we render each source and target mesh from the same 10 fixed viewpoints and compare the corresponding views against each other. To ensure balanced evaluation across our dataset, we first compute the average score within each object category, then average these category scores to obtain the final metric values. 

\paragraph{Human Evaluation}
To complement these quantitative metrics, we also conduct a user study on Amazon Mechanical Turk (AMT) with 15 participants to evaluate our results. Each participant viewed a source object and two versions of the textured target mesh from different methods, selecting which result better preserves the texture and the appearance of the reference object. Each comparison received ratings from two independent evaluators.

\subsection{Results}

We provide qualitative examples, qualitative and quantitative comparisons with baselines, and an ablation study with qualitative and quantitative analysis.

\vspace{-8pt}
\paragraph{Qualitative Examples} 
Figure \ref{fig:packs} provides four examples illustrating how \ourmethod{} transfers texture from the source shape (left column) to three other shapes. Notice how colors match the semantics of the location, despite significant shape variations. for example the eyes of the fish at the bottom row. Figure~\ref{fig:results} shows additional results along with the corresponding target geometries.

\vspace{-8pt}
\paragraph{Quantitative Comparisons:}
Table \ref{tab:performance_comparison} compares our \ourmethod{} approach with all baseline methods. We find that \ourmethod{} outperforms baselines in both evaluation metrics, successfully capturing the appearance of the original objects.  We also report the inference time for texturing a new mesh for each method.

\vspace{-8pt}
\paragraph{Qualitative Comparisons:} 
We show qualitative comparisons \hspace{4pt}between our \hspace{4pt}approach and the\hspace{4pt} baselines in Fig.~\ref{fig:comparisons}. \hspace{4pt} As demonstrated, MVEdit tends to deviate significantly from the input image, only taking vague inspiration from the reference. TEXTure and EASI-TEX produce artifacts due to their iterative painting techniques.

\begin{figure}[!htbp]
    \centering
    \includegraphics[width=\columnwidth]{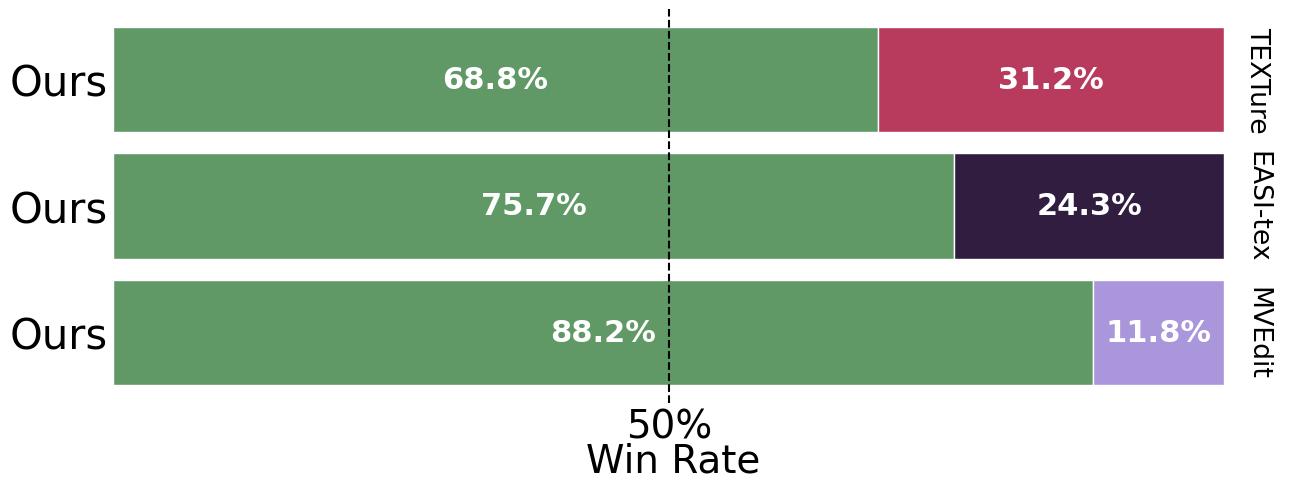}  %
    \caption{Results of human evaluation study. \ourmethod{} Was compared with three baseline approaches in a 2-alternative forced choice setting over $\sim 750$ questions. Raters strongly favored \ourmethod{} for better transfer of appearance. }
    \label{fig:human_study}
\end{figure}

\paragraph{User Study:} 
We conducted a user study to evaluate the quality of \ourmethod{} compared with baselines. For each baseline, we presented the source object and two target shapes, one textured with \ourmethod{ } and one with the baseline, in a 2-alternative forced choice (2AFC) setting.  We used Amazon Mechanical Turk (AMT) and paid raters above the minimum wage. Figure  \ref{fig:human_study} shows the results of these comparisons. \ourmethod{} was strongly preferred by raters over all three competing approaches.

\subsection{Ablation Studies}

\begin{figure}[!htbp]
    \centering
    \setlength{\tabcolsep}{1pt}
    \newlength{\imgwidth}
    \setlength{\imgwidth}{0.26\linewidth}  %
    {\small
    \begin{tabular}{ccc}
        \includegraphics[width=\imgwidth, trim=60 60 20 50, clip]{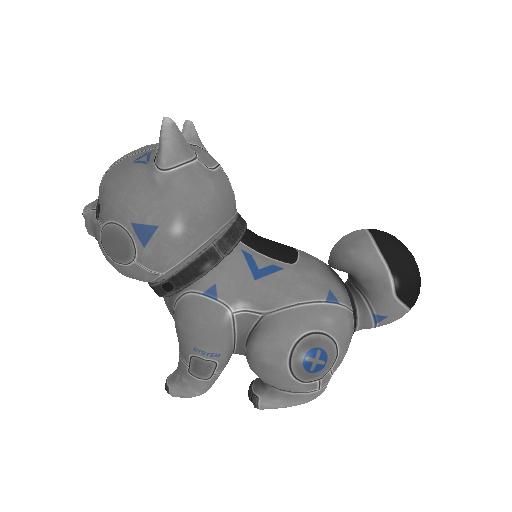} &
        \includegraphics[width=\imgwidth, trim=60 60 20 50, clip]{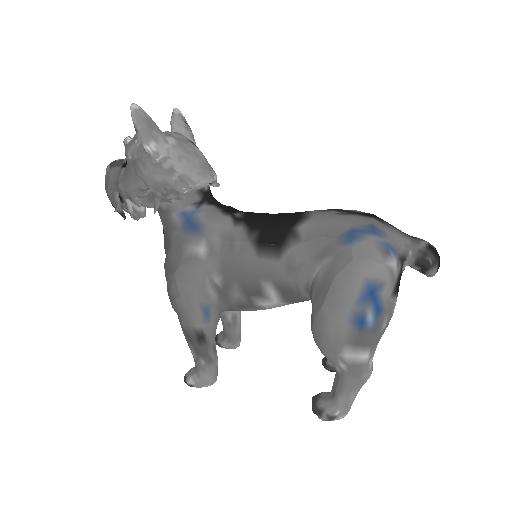} &
        \includegraphics[width=\imgwidth, trim=60 60 20 50, clip]{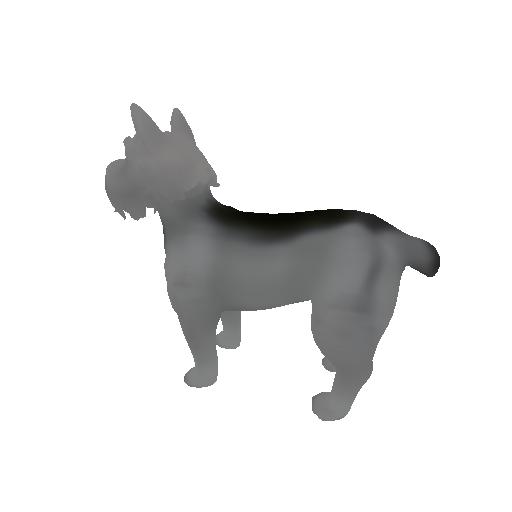} \\
        source shape & w/o network & w/o \( \mathcal{L}_{\text{app}} \) \\
        \includegraphics[width=\imgwidth, trim=60 60 20 105, clip]{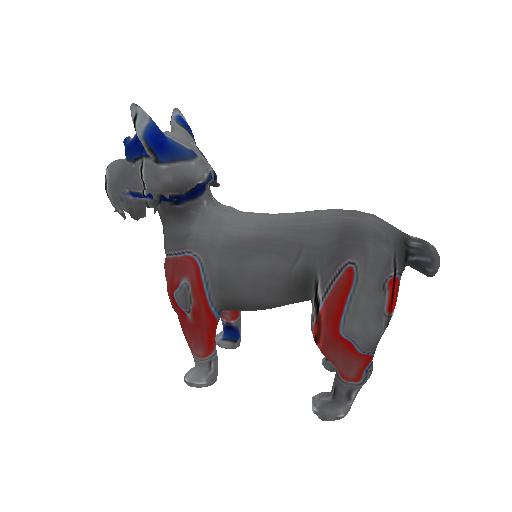} &
        \includegraphics[width=\imgwidth, trim=60 60 20 50, clip]{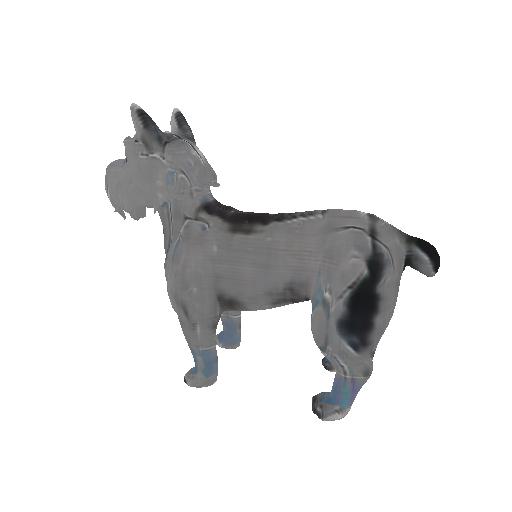} &
        \includegraphics[width=\imgwidth, trim=60 60 20 50, clip]{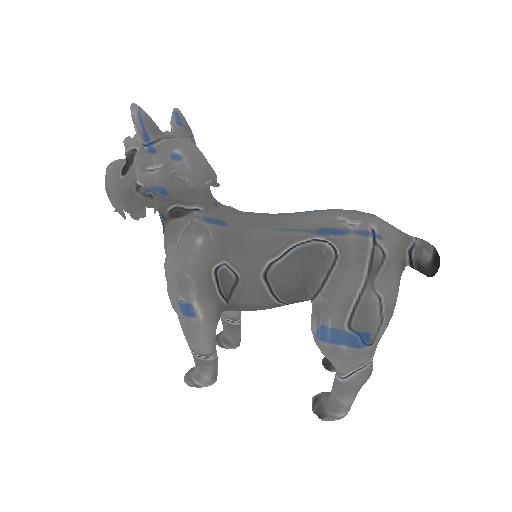} \\
        w/o \( \mathcal{L}_{\text{MSE}} \) & w/o augmentations & ours \\
    \end{tabular}
    }
    \caption{\textbf{Qualitative Ablation Study.} We demonstrate the effect of removing each component of our pipeline on a single example. As shown, each component is essential for successful texture transfer.
    }

    \label{fig:ablation}
\end{figure}

We conducted an ablation study to evaluate the contribution of each component in our method. The quantitative and \hspace{2pt}quantitative\hspace{2pt} results\hspace{2pt} presented in Table~\ref{tab:ablation}\hspace{2pt} and Fig.~\ref{fig:ablation}\hspace{4pt} respectively, demonstrate that each component is essential for optimal performance. Our analysis reveals that excluding  \( \mathcal{L}_{\text{mse}} \)  leads to color predictions that deviate significantly from the original texture. Similarly, removing \( \mathcal{L}_{\text{app}} \) results in blurry outputs that fail to capture complex texture patterns.  We further found that omitting data augmentation causes the network to overfit to specific triplane projections, thereby limiting the method's generalization capabilities.

\begin{figure}[!htbp]
    \vspace{-16pt}
    \centering
    \setlength{\tabcolsep}{1pt}
    {\small
    \begin{tabular}{ccc}
        \multicolumn{3}{c}{} \\
        \includegraphics[width=0.32\linewidth, trim=60 60 20 50, clip]{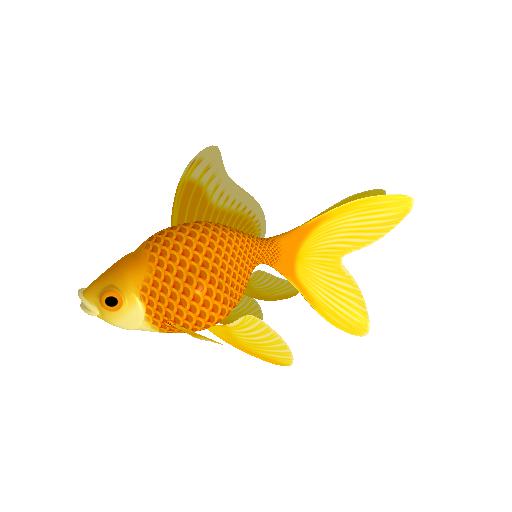} &
        \includegraphics[width=0.32\linewidth, trim=60 60 20 50, clip]{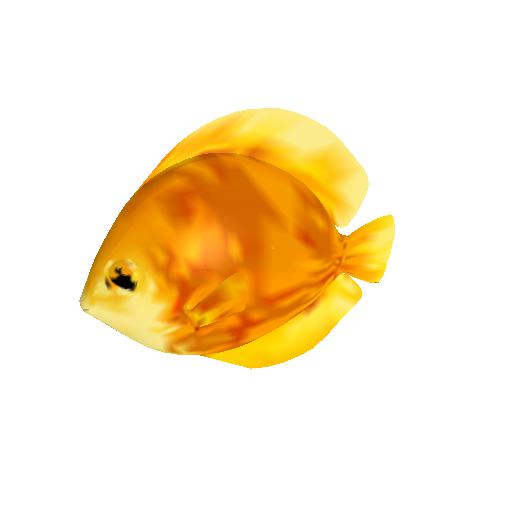} &
        \includegraphics[width=0.32\linewidth, trim=60 60 20 50, clip]{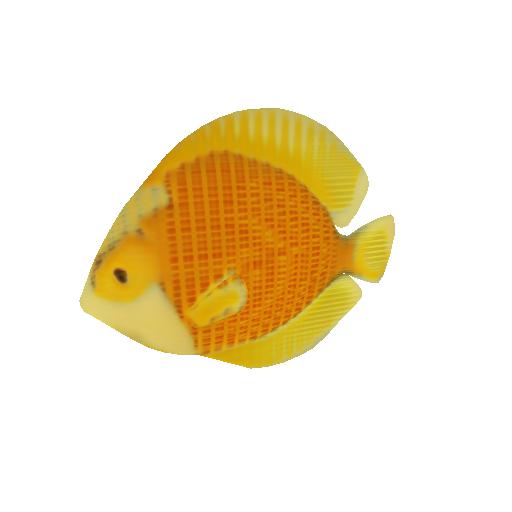} \\
        \includegraphics[width=0.32\linewidth, trim=60 60 20 105, clip]{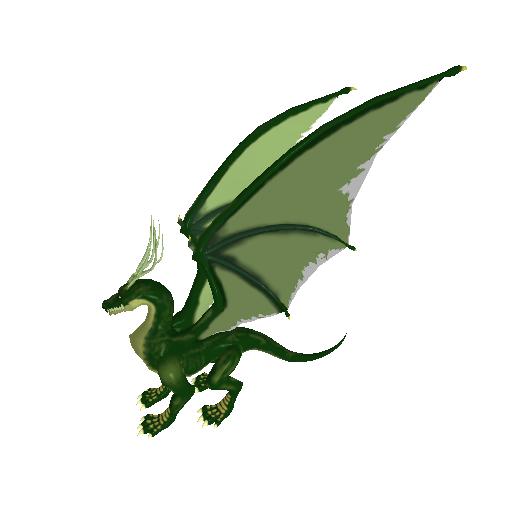} &
        \includegraphics[width=0.32\linewidth, trim=60 60 20 50, clip]{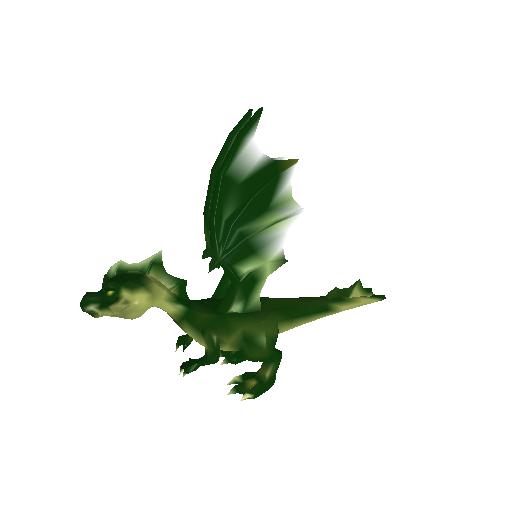} &
        \includegraphics[width=0.32\linewidth, trim=60 60 20 50, clip]{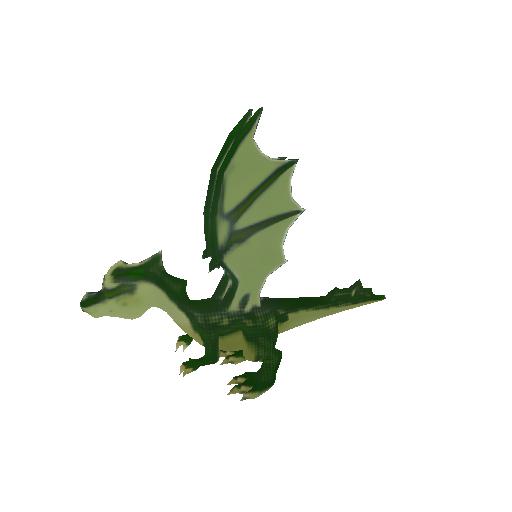} \\
        source shape & w/o network & ours \\
    \end{tabular}
    }
    \caption{ 
    \textbf{Network Ablation.} A comparison between our full method (right) and the nearest neighboring baseline (middle), which replaces the neural network in our pipeline and transfers color based on the closest matching neighbors. As seen, this baseline fails to transfer complex texture details, unlike our full method.
    }
    \label{fig:no network}
\end{figure}

We also ablate the contribution of the neural network in our pipeline. In the "w/o network" baseline, we utilize pre-extracted DIFF3f semantic features and identify the nearest neighbor for each target mesh feature within the source mesh features. Using these nearest-neighbor matches, we transfer colors from the source mesh to the target vertices and then render the target mesh with its new colors. The impact of this simplification is presented in Table~\ref{tab:ablation} and Fig.\ref{fig:ablation}, demonstrating the degraded quality of this approach in transferring the texture details compared to our full pipeline. Two additional examples are shown in Fig.\ref{fig:no network}.

\begin{table}[t]
    \caption{\textbf{Quantitative Ablation Study.}
We demonstrate the effect of removing component of \ourmethod{} with quantitative metrics averaged across our evaluation set. The results show that each component is essential for achieving successful texture transfer.}
    \label{tab:ablation}
    \centering  
    {\small
    \begin{tabular}{lcc}  
        \toprule
        \textbf{Method} & \textbf{SIFID $\downarrow$} & \textbf{CLIP sim. $\uparrow$} \\ 
        \midrule
        w/o network & 0.23 & 0.86  \\
        w/o  \( \mathcal{L}_{\text{MSE}} \) & 0.23 & 0.86  \\
        w/o \( \mathcal{L}_{\text{app}} \)& 0.28 & 0.85  \\
        w/o augmentations & \textbf{0.21} & 0.85  \\
        TriTex (ours) &  0.22 & \textbf{0.87}\\
        \bottomrule
    \end{tabular} 
    }
\end{table}

\section{Discussion}

\begin{figure}[h!]
    \centering
    \setlength{\tabcolsep}{1pt}
    \setlength{\imgwidth}{0.32\linewidth}  
    {\small
    \begin{tabular}{ccc}
        \multicolumn{3}{c}{} \\
        \includegraphics[width=\imgwidth, trim=60 60 20 50, clip]{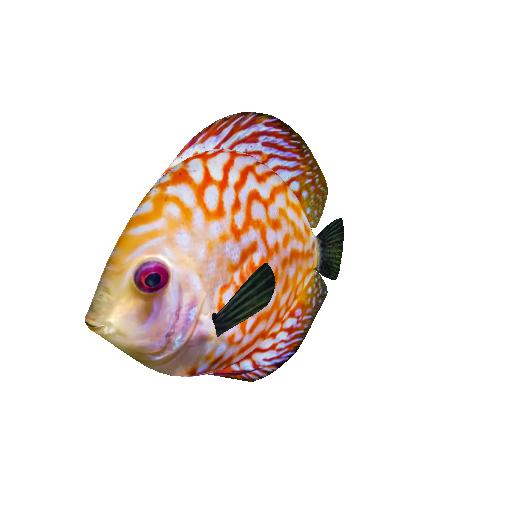} &
        \includegraphics[width=\imgwidth, trim=60 60 20 50, clip]{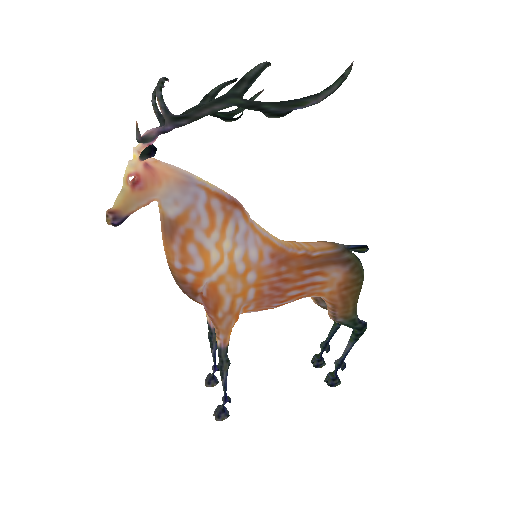} &
        \includegraphics[width=\imgwidth, trim=60 60 20 50, clip]{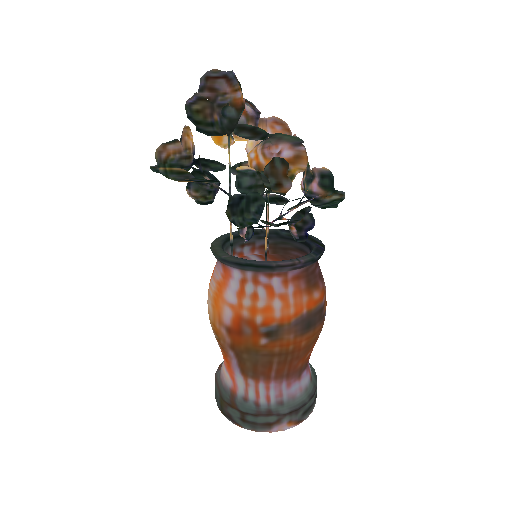} \\
        Source Shape & Output 1 & Output 2 \\
    \end{tabular}
    }
    \caption{\textbf{Cross Category Texture Transfer.} Texture transfer between objects with partial semantic correspondence (fish and deer) and versus objects with  minimal semantic similarity (fish and vase with flowers).}
    \label{fig:cross_category}
\end{figure}

\paragraph{Cross-Category Object Transfer}
Although our methods is desinged to transfer appearance between objects from the same category, it is interesting to examine the results when transferring appearance between objects with varying degrees of semantic similarity.
As shown in Figure \ref{fig:cross_category}, when transferring between objects with shared semantic structure (e.g. parts with similar functionality or spatial relationships), \ourmethod\ preserves meaningful texture patterns aligned with these semantic features. However, when transferring between objects with little semantic overlap, the resulting texture patterns become more stochastic, as the semantic mapping becomes less well-defined.

\begin{figure}[!htbp]
    \centering
    \setlength{\tabcolsep}{1pt}
    {\small
    \begin{tabular}{cccc}
        {\includegraphics[width=0.22\linewidth, trim=60 60 60 60, clip]{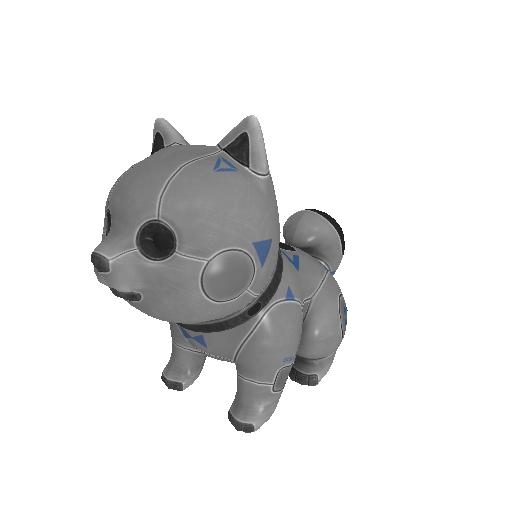}} &
        {\includegraphics[width=0.22\linewidth, trim=30 30 30 30, clip]{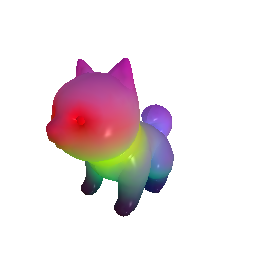}} &
        {\includegraphics[width=0.22\linewidth,trim=100 80 60 100, clip]{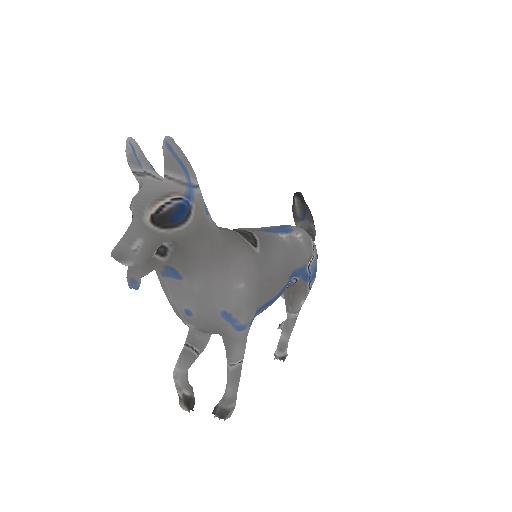}} &
        {\includegraphics[width=0.22\linewidth, trim=50 40 30 50, clip]{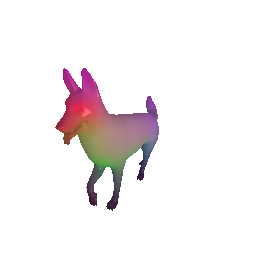}}
        \\[0pt]
        {\includegraphics[width=0.22\linewidth,trim=50 100 10 100, clip]{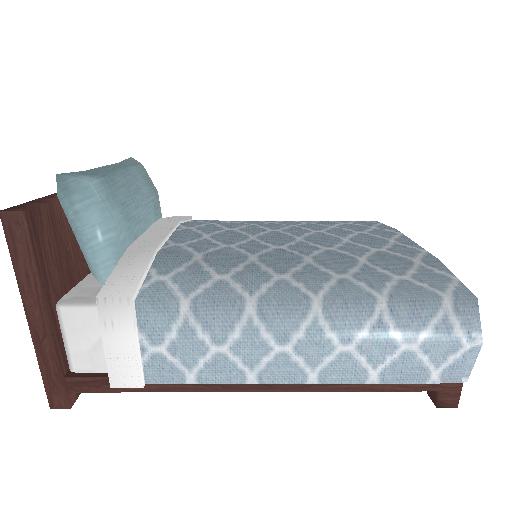}} &
        {\includegraphics[width=0.22\linewidth,trim=25 50 5 50, clip]{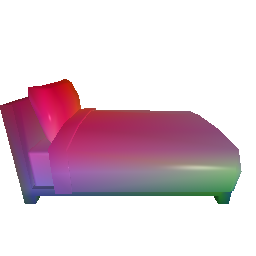}} &
        {\includegraphics[width=0.22\linewidth,trim=50 120 20 100, clip]{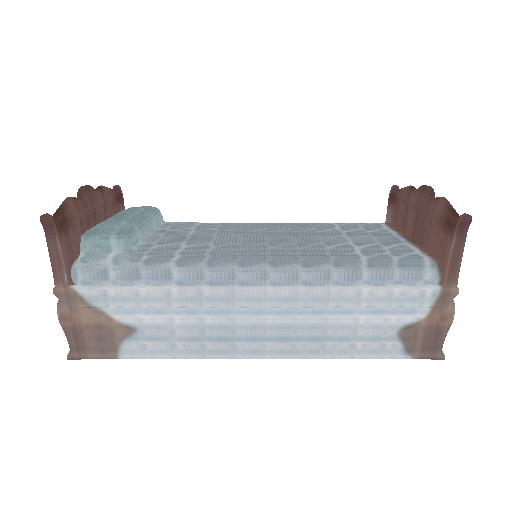}} &
        {\includegraphics[width=0.22\linewidth,trim=25 60 10 50, clip]{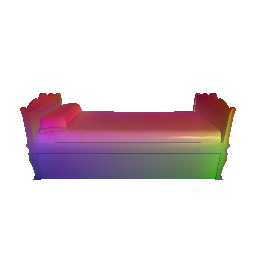}}
        \\[0pt]
        source shape & source shape PCA & target shape & target shape PCA
    \end{tabular}
    }
    \caption{%
        \textbf{Limitations.} (Top) Since our network lacks a generative prior and the source dog has no visible tongue, it cannot synthesize an appropriate color for the target dog's tongue.
        (Bottom) The network incorrectly transfers the source bed's sheet texture to the target bed frame, as PCA analysis shows their feature representations are similar. 
    }
    \label{fig:limitations}
\end{figure}

\vspace{-12pt}
\paragraph{Limitations}
Despite the effectiveness of our approach, it has several inherent limitations. First, our method does not possess generative capabilities, which means it cannot synthesize novel details that are not present in the source texture. This limitation becomes particularly apparent when there is significant geometric variation between the source and target meshes, as our method can only map existing texture patterns rather than create new ones to accommodate the structural differences (Figure \ref{fig:limitations}, top). Second, without leveraging priors from text-to-image models, our approach lacks the ability to compensate for cases where the semantic features fail to capture sufficient detail or establish accurate correspondences. This can result in less detailed or less coherent texture transfers in regions where the feature matching is ambiguous or imprecise (Figure \ref{fig:limitations}, bottom). Additionally, our method expects the target shape to be at the same orientation as the source shape. When enabling arbitrary rotations during training, the model becomes invariant to the object's orientation, but the ability to reproduce fine details such as eyes or wheels is reduced.

\vspace{-12pt}

\section{Conclusions}

We have presented a method that allows transferring a texture learned from a single exemplar to similar shapes. The learned texture is challenging in the sense that it is a non-stationary texture, where its patterns can be strongly correlated with the semantics of the shapes. 
Mapping such semantic textures on a target shape, necessarily requires, at least implicitly, extracting its semantics. For that, our model uses pre-trained 3D lifted semantic features.

Our key challenge was to process the semantic features in a way that allows generalization from a single instance, in order to perform the texture learning with minimal supervision. To achieve this, we re-rendered them into a triplane representation, which preserves feature proximity and connectivity, allowing for efficient processing and mapping to the corresponding colors. We introduced an effective training method using a single mesh with appearance losses while applying augmentations.  The limitation of our technique is that it currently does not possess generative abilities and cannot create details that do not appear in the source texture.

In future work, we aim to extend this feed-forward approach for fast texturing by training the model to predict the mesh texture from a reference image, using cross-attention layers between the generated triplane and the reference image. Additionally, we plan to leverage pre-trained generative models to enhance the mapped textures, increasing both resolution and quality with their learned priors.

\section*{Acknowledgements}
We thank Rinon Gal, Yoad Tewel and Or Patashnik for their constructive suggestions and insightful comments.
{
    \small
    \bibliographystyle{ieeenat_fullname}
    \bibliography{main}
}

\setcounter{section}{0}
\setcounter{table}{0}
\setcounter{figure}{0}
\maketitlesupplementary

\section{Dataset}
Tables \ref{tab:objects_per_category} and \ref{tab:objects_list} show our dataset composition, including the number of objects per category and their corresponding Objaverse IDs.
\begin{table}[h]
\caption{Object Categories and Counts}
\centering
\begin{tabular}{|l|c|}
\hline
Category & Number of Objects \\
\hline
Animal & 10 \\
Bed & 4 \\
Bird & 7 \\
Dragon & 5 \\
Character & 5 \\
Fish & 6 \\
Guitar & 4 \\
Plant & 4 \\
Vase with flowers & 7 \\
\hline
\end{tabular}
\label{tab:objects_per_category}
\end{table}

\section{Implementation Details}
Our method consists of three main components: the semantic feature projection module, the triplane processing network, and the coloring MLP.

For the semantic features, we utilize Diff3F with Stable Diffusion v1.5 and ControlNet v1.1. We render depth and normal maps at 512×512 resolution from 16 viewpoints uniformly distributed on a sphere. The diffusion process uses 30 denoising steps with a guidance scale of 7. The extracted features, which combine information from the UNet and DINO features resulting in features of dimension 32×32×2048 per view, and are then aggregated per vertex.

The feature projection module generates a triplane of size 256×256. Each feature plane is created by concatenating features from two opposite orthographic projections. To compensate for the relatively low spatial resolution, we incorporate positional encoding into the input.

The triplane processing network consists of 6 residual ConvNets block which reduces the channel dimension to 64,  followed by  triplane-aware UNet from ~\cite{Sin3DM} which output features of dimension 256x256x12.  The coloring MLP is a lightweight network consisting of two linear layers. 

For training, we employ the Adam optimizer with learning rates of 1e-2 and 1e-3 for the triplane processing network and coloring MLP, respectively. We sample 30 random camera views per iteration at 256×256 resolution. The preprocessing augmentation generates 5 variants of the input mesh through combinations of scaling $(0.5-1.7)$ and rotations ($\pm15^\circ$) around each axis. During training, we apply similar augmentations and add random translations (±0.1). Training takes approximately 1 hour on a single A100 GPU.

\section{User Study Details}
\begin{figure}[h!]
\centering
\begin{subfigure}{\linewidth}
    \includegraphics[width=\linewidth]{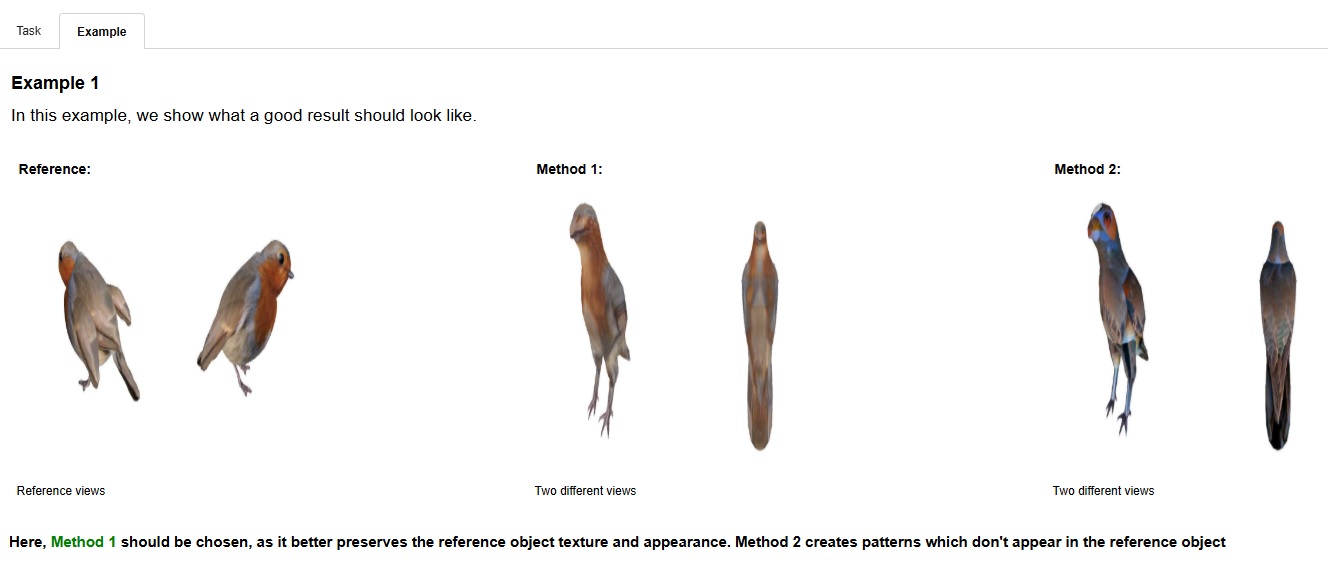}
    \caption{Example comparison presented to participants, showing source object and two target results for selection.}
\end{subfigure}
\vspace{0.2cm}
\begin{subfigure}{\linewidth}
    \includegraphics[width=\linewidth]{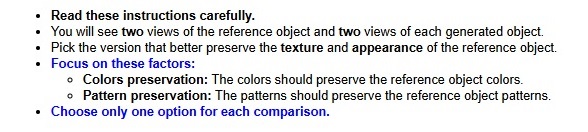}
    \caption{Task instructions detailing evaluation criteria and selection guidelines.}
\end{subfigure}
\caption{User study interface and instructions. Participants were asked to select which result better preserves the style of the source object while adapting to the target shape.}
\label{fig:user_study_details}
\end{figure}

We conducted a two-alternative forced choice (2AFC) study on Amazon Mechanical Turk to assess texture transfer quality. Figure \ref{fig:user_study_details} shows the evaluation interface and task instructions provided to participants.  Each task displayed a source object and two textured target shapes for comparison.

\begin{table}[!htbp]
\centering
\small 

\caption{Full list of objects with their Objaverse IDs} 
\begin{tabular}{|p{2.25cm}|p{5.15cm}|}
\hline
\multicolumn{1}{|c|}{\textbf{Category}} & \multicolumn{1}{c|}{\textbf{Objaverse ID}} \\
\hline
Bird & a268cb1c8e3c4b328a4a797632805a22 8b99562d27d84a29bfad2c33306bd172 80cc44761a294682bd998b5b17287c8c b8c3b9076fd14b0e934f2784d8de105a 32a49fbded87487383f875b7f8998fc2 234a8576b3d0409aab8545c72ba7e1db e05d043c884d4c5bb916e4c43871750a \\
\hline
Fish & 7de2969ef2ce44578746588729f19459 9945e1eb5a6247cf9623506025d92e7b 793c85d819f140c29d14a5dc424c128a 551d23edef9c4a78b67b6bba9e8f6294 f42aa80e36a44ccab242aa6868b3b5c2 74e57a9de7a24975b02d236ea3be614f \\
\hline
Vase with flowers & 9ea304aab8b345e5839eb31d4d88e157 39db0a1edb6449ee98cf3cb64afb72c1 cbafed33e2f7412c97ba3941c399b2df 647a28ca37a84e0bbc312d0b8044452d f011d24ce98a4de49dbb68a2472a8580 4f1403f9b68441daa824179c9f62c53a f5241a92db634dfba7c237fe47bc909b \\
\hline
Bed & b19855811635449288827767b45d4b38 952e4e69261b4f419d1a7f7e9df955dc 210be84bcb5449f5a9f66a923c8ae307 7b13b36ba2304912afc9840caea731c6 \\
\hline
Guitar & 2007af7561fe46958d1f7e92dff8a40d 4dd67b2cea5143e7b56450629f8cb120 a44930f9c14544a6ae6967d5544417e8 0f4e0e54644e4fa1b96eaf033e17db6f \\
\hline
Dragon & 31a959e19e85458488d2ebff9ecb9793 62b512c628da4b16ba4a82cb0acdbc62 942c32cd4f8b4f028aee817a3c7947d9 268a461b9fae45adb20e0e208a0861a4 ff88bccf5b2c4b0fa1fb1ee19c80d5a4 \\
\hline
Animal & 6a0a93cf64ef4cfdb06fef0641286a06 cd236d93f16346c580bbf9f0b03d0e14 7d8266ae0e764478a03c2477dde6629f 977f9efc21084dfab70a9ed35f66873b 64998ee900d641d2b5096caaa5cdf006 f653fb955a4848e99c29b7da1e0a0a42 b4ce5dbdc0da4c72a6d00c06ec8db662 6ff3ec85501e444db8d0161c0dcfaedb c67d1a28ed8f4069916d9f6d999590f9 b4215f3c452c4e7cbe845b56251d2877 \\
\hline
Character & e1502d8f865f451c8022e7164521c22b bc851b9f608146f193b8a3dc78506b9f 88ed6191446749b9a9e24b995bcb5e1d 5aa0a3cfcd4b44bba5a992a14238619a e93f7209713c442390ca9bb959caee3d \\
\hline
Plant & 6f171aa67ad4434895366886abd02dbb 54c7520ace8446e89daad21ea03d7dca 1bb0cf7261174670ad1134093875e1d1 57972124483145b4a4bbf4fd4caca6e7 \\
\hline
\end{tabular}
\label{tab:objects_list}
\end{table}

\end{document}